\documentclass[prl,twocolumn,showpacs,preprintnumbers,amsmath,amssymb]{revtex4}
\usepackage{graphicx}
\usepackage{dcolumn}
\usepackage{bm}
\usepackage{textcomp}

\begin{document}

\title{Diffraction of a Bose-Einstein condensate from a Magnetic Lattice on a Micro Chip}

\author{A. G\"unther}
\homepage{http://www.pit.physik.uni-tuebingen.de/zimmermann/}
\email{aguenth@pit.physik.uni-tuebingen.de}
\author{S. Kraft}
\author{M. Kemmler}
\author{D. Koelle}
\author{R. Kleiner}
\author{C. Zimmermann}
\author{J. Fort\'agh}
\affiliation{Physikalisches Institut der Universit\"at T\"ubingen,
Auf der Morgenstelle 14, 72076 T\"ubingen, Germany}

\date{\today}

\begin{abstract}
We experimentally study the diffraction of a Bose-Einstein condensate from a magnetic lattice, realized by a set of 372 parallel gold conductors which are micro fabricated on a silicon substrate. The conductors generate a periodic potential for the atoms with a lattice constant of $4\mu\text{m}$. After exposing the condensate to the lattice for several milliseconds we observe diffraction up to $5^\text{th}$ order by standard time of flight imaging techniques. The experimental data can be quantitatively interpreted with a simple phase imprinting model. The demonstrated diffraction grating offers promising perspectives for the construction of an integrated atom interferometer. 
\end{abstract}


\maketitle
With magnetic fields of miniaturized current conductors, ensembles of ultra cold atoms can be trapped and manipulated on a spatial scale of micrometers and below \cite{Folman2002}. Typically, the conductors are micro fabricated on a chip and form wave guide type potentials in which the atoms are trapped close to the surface of the substrate. Since also Bose-Einstein condensates can been loaded into such micro traps \cite{Ott2001,Hansel2001}, the possibility of matter wave interference is intensely discussed, with the future perspective of integrated atom optics for sensitive interferometric detection of forces with a high spatial resolution.

Up to now, in purely magnetic micro traps no experiments have been carried out which are sensitive to the phase coherence of the condensate wave function. As a step in this direction standard optical lattice potentials have been combined with miniaturized magnetic traps \cite{Wang2005}, however, in magnetic micro chips alone no interference effects have yet been observed. This is because the design and construction of a suitable magnetic structure on a chip is not obvious. Pronounced diffraction from a periodic potential, for instance, is expected only for small lattice constants on the scale of micro meters and below. This requires to bring the condensate very close to the current conductors that generate the potential. At close distances to a metallic surface strong losses of atoms have been observed due to thermal magnetic field fluctuations \cite{Henkel1999a,Jones2004,Lin2004}. Also geometric imperfections of the current conductors introduce perturbations of the trapping potential which are strongest in the direct vicinity of the conductor \cite{Esteve2004a}.

In this work we present an experiment that avoids these difficulties and allows for the first observation of atomic matter wave interference in a magnetic micro trap. With a magnetic lattice we imprint a periodic phase pattern onto the macroscopic wave function of the condensate and subsequently observe its temporal evolution.  Diffraction peaks up to $5^\text{th}$ order can be observed. Similar phase imprinting methods are used in experiments for generating vortices and solitons in Bose-Einstein condensates \cite{Matthews1999,Burger1999}. In the context of atom optics phase imprinting is comparable to atom diffraction from matter gratings in the Raman Nath regime \cite{Henkel1994}, with the additional aspect that, for a condensate, the atomic interaction energy may play a role.  While thermal atoms have previously been scattered from magnetic micro structures \cite{Rosenbusch2000} we here present diffraction of coherent atomic matter waves.

\begin{figure}
\centerline{\scalebox{0.6}{\includegraphics{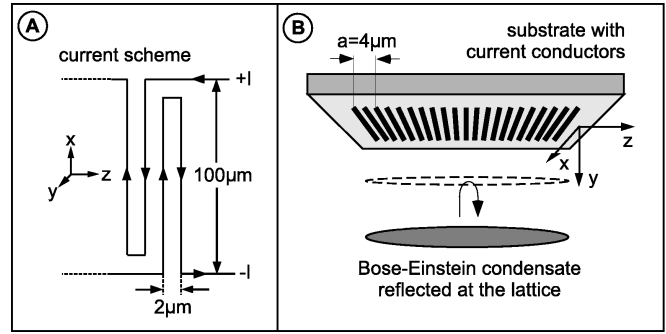}}}
\caption{Sketch of the experimental situation. A) Current scheme of the magnetic lattice. The currents in neighbouring conductors are equal and oppositely poled. B) The condensate approaches the lattice during a vertical oscillation (y-direction) in an elongated harmonic trap. After phase imprinting the condensate is released from the trapping potential.} \label{fig:Figure1}
\end{figure}

The experimental situation is sketched in Figure \ref{fig:Figure1}.
A Bose-Einstein condensate consisting of $1.2\cdot 10^5$ $^{87}\text{Rb}$ atoms in the F=2, m$_{\text{F}}$=2 hyperfine state is trapped in an elongated magnetic trap close to the surface of a chip that carries a magnetic lattice \cite{Guenther2005a}. It consists of a set of 372 parallel conductors perpendicular to the long axes of the trap, each $1\mu\text{m}$ wide and separated by $1\mu\text{m}$ gaps \cite{comment1}. The current of $I=0.2\mbox{mA}$ in each conductor is oppositely oriented for neighboring wires. This results in a periodic potential for the atoms with a sinusoidal modulation along the axial direction (long axis of the trap) and an exponential decay in the vertical direction (perpendicular to the chip surface) \cite{Opat1992,Lau1999}. The condensate approaches the lattice during a controlled vertical oscillation inside the trap (period T=13.2 ms) and interacts with the lattice only for a short time at the turning point of the oscillation. During this time the lattice potential imprints a phase onto the macroscopic condensate wave function. After being reflected from the lattice, the condensate is released for ballistic expansion by turning of all magnetic fields. The axial gradient of the imprinted phase results in an axial velocity distribution that can be monitored by absorption imaging after some time of flight. The strength of phase imprinting is controlled by varying the amplitude of the vertical oscillation. By using this kind of pulsed interaction, surface induced losses and decoherence effects \cite{Henkel1999a,Jones2004,Lin2004} are minimized because of short interaction time.

The lattice constant $a=4\mu\text{m}$ is one order of magnitude larger than in typical experiments with optical lattices. Thus, the energy due to the repulsive atomic interaction significantly exceeds the effective "recoil energy" of the lattice $E_r=\frac{1}{2}m v_l^2$ with $m$ being the mass of the atoms and $v_l =2\pi\hbar/ma=1.14mm/s$ the reciprocal lattice velocity. Thus, the atomic interaction can not be neglected, a priori. The experimental observation, however, shows that the interaction comes into play primarily during the ballistic expansion, after phase imprinting.

\begin{figure}
\centerline{\scalebox{0.6}{\includegraphics{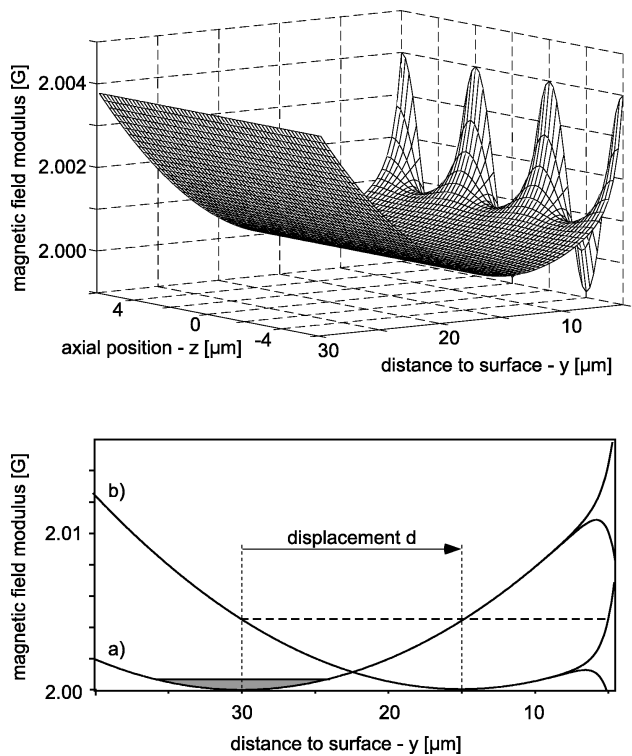}}}
\caption{Shape of the lattice potential (to scale). Close to the surface (small values of y) the harmonic trapping potential is distorted by the periodic magnetic field of the lattice. The lower figure shows vertical cuts of the potential. Initially the condensate is generated in a harmonic trap at a distance of $30\mu\text{m}$ from the surface (a).  Then the minimum is suddenly shifted closer to the surface by a controlled displacement $d$. In curve b) $d$ amounts to a value of $15\mu\text{m}$. The dashed line indicates the potential energy of the condensate after the displacement.} \label{fig:Figure2}
\end{figure}

The experimental setup is similar as in our previous experiments \cite{Fortagh2003}. Typically, $3\cdot 10^8$ Rubidium atoms from a magnetooptical trap are transferred into a standard magnetic trap and precooled by forced evaporation to a temperature of $5\mu\text{K}$. Next, the atoms are adiabatically shifted into a micro trap transport system that allows to generate a harmonic trapping potential at an arbitrary position within a volume of \mbox{1.5 x 0.3 x 20 $\text{mm}^3$} (x, y, z - direction) \cite{Guenther2005a}. The transport system consists of an assembly of $100\mu\text{m}$ wide gold conductors electroplated on both sides of a $250\mu\text{m}$ thick substrate (``carrier chip''). Since the atoms are held at a large distance from the surface of this ``carrier chip'', fragmentation effects \cite{Kraft2002} can be avoided. The magnetic trap used for the present experiment is characterized by the axial and the radial oscillations frequencies of $2\pi\cdot 16 \text{Hz}$ and $2\pi\cdot 76 \text{Hz}$, respectively. The micro chip with the lattice conductors (Fig. \ref{fig:Figure1}) is mounted onto the surface of the carrier chip. The magnetic field of the lattice adds to the field of the wave guide trapping potential resulting in a total potential as shown in Figure \ref{fig:Figure2}.
It features a parabolic wave guide potential oriented along the z-direction and a periodic structure with a lattice constant of $4\mu\text{m}$ in the vicinity of the conductors. The periodic structure reveals a steep potential increase near every second conductor. In between, the potential is lowered by the field of the nearby conductors, leading to saddle points and limiting the trap depth. Atoms that pass the saddle points fall into steep quadrupole traps which emerge close to the conductor. Such atoms are partially lost due to Majorana spin flips or surface induced losses.

\begin{figure}[b]
\centerline{\scalebox{0.55}{\includegraphics{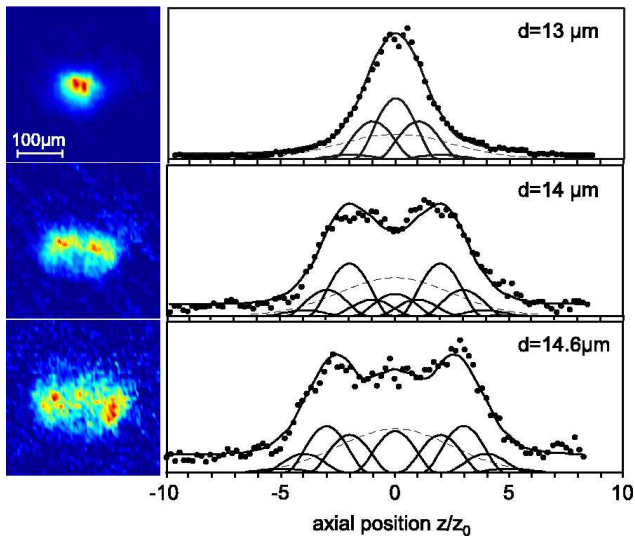}}}
\caption{(Color online) Absorption images and axial density profiles after $\tau=20$ ms time of ballistic expansion for three different displacements $d$ of $13\mu\text{m}$, $14\mu\text{m}$ and $14.6\mu\text{m}$. The color code represents the density distribution (blue=low density, red=high density). The distance $z_0=v_l\cdot \tau$ corresponds to one reciprocal lattice velocity $v_l$. The density profiles can be described by the sum of up to 5 overlapping diffraction orders (solid line). While the relative strength of each diffraction order is given by the phase imprint parameter $S$ an additional gaussian density distribution (dashed line) is used to take account the fraction of thermal atoms.} 
\label{fig:Figure3}
\end{figure}

Initially a condensate with $1.2\cdot 10^5$ atoms is generated inside a trap, which is placed at a distance of $30\mu\text{m}$ from the lattice conductors. The trap minimum is then suddenly shifted towards the lattice by a variable displacement $d$ (Fig. \ref{fig:Figure2}) \cite{comment2}. After shifting the potential towards the surface, the condensate starts a vertical oscillation. It is partially reflected from the lattice potential and swings back to the starting position. 12ms after the initial displacement, all magnetic fields are suddenly switched off and the condensate is released for ballistic expansion. The diffraction of the condensate is detected after 20ms time of flight by standard absorption imaging, with a spatial resolution of about $5\mu\text{m}$. For a quantitative analysis of the diffraction process, images have been taken for 24 different displacements $d$ ranging from 0 to $15.2\mu\text{m}$. Each image is integrated along the vertical direction resulting in an axial line density profile. Figure \ref{fig:Figure3} shows typical results for three different displacements $d$ of $13\mu\text{m}$, $14\mu\text{m}$ and $14.6\mu\text{m}$. 
For displacements up to $11\mu\text{m}$ more than 90\% of the atoms are reflected. For larger displacements the reflectivity rapidly decreases to a value of about 10\% at $d=15\mu\text{m}$. This is consistent with the onset of losses beyond the saddle point.

A complete theoretical description of the diffraction process would require a three dimensional numerical simulation of the Gross-Pitaevskii-equation including the initial oscillation, the ballistic expansion, and the losses at the lattice. This rather involved program is beyond the scope of the present work. However, it is possible to quantitatively describe the essential features of the density profiles with a simplified analysis (solid lines in Fig. \ref{fig:Figure3}). It follows the lines of previous theoretical work on diffraction of single atoms form a periodic optical potential generated by an evanescent light wave \cite{Henkel1994}. Since the condensate is exposed to the lattice potential only for a short time ($< 1\text{ms}$), the lattice primarily changes the local phase of the condensate wave function. This is equivalent to diffraction in the limit of a 
thin lattice (Raman Nath regime). We thus assume that the details of phase imprinting can be summarized by a phase function $\phi(z,d)=S(d)\cos{kz}$ which contains the potential modulation with a lattice vector $k=2\pi/a$ and a dimensionless phase imprint parameter $S(d)$ that describes the strength of phase imprinting. The condensate wave function directly after phase imprinting then reads:
\[
\Psi(x,y,z)=\sqrt{n(x,y,z)}e^{-iS(d)\cos{kz}}
\]
with $n(x,y,z)$ being the density distribution of the condensate. The relative number of atoms in the different diffraction orders is now obtained by expanding the wave function as a sum of momentum eigenfunctions of the axial motion \cite{Henkel1994}. By exploiting the properties of the Bessel-functions of the first kind $J_n$ one finds:
\[
\Psi(x,y,z)=\sqrt{n(x,y,z)}\sum_{n} (-i)^n J_n\left(S\left(d\right)\right)e^{inkz}
\]
Obviously, $\Psi(x,y,z)$ consists of a discrete superposition of momentum eigenfunctions with wave vectors $k_n=nk$. The probability for an atom to be diffracted in the $\text{n}^\text{th}$ order is thus proportional to $\left|J_n\left(S\left(d\right)\right)\right|^2$. 

As the images were taken after 20ms of ballistic expansion the measured data reveal the momentum distribution of the diffracted condensate. Therefore the axial density profiles can be compared to a model function which is composed of 11 inverted parabolas: one for the $0^\text{th}$ order and a symmetric pair for each higher order (up to $5^\text{th}$ order). The parabolic shape which is expected for a single condensate after free ballistic expansion \cite{Castin1996} is here taken as the model function $\rho_0(z)$ for the shape of each individual diffraction peak. The positions of the parabolas are equally spaced by the expected separation of adjacent diffraction orders. This separation is given by $v_l \tau$ with $\tau=20ms$ being the time of ballistic expansion. Following the above argument the relative amplitude of the different diffraction orders is proportional to $\left|J_n(S)\right|^2$. The model function for the condensate line density can therefore be written as:
\[
\rho_{\text{cond}}(z)=A \sum_{n=-5}^{+5}\rho_0(z-n v_l \tau)\left| J_n(S)\right|^2
\]
Besides the overall amplitude $A$ and an overall offset in the density, the only free adjustable parameter is the phase imprint parameter $S$. Taking into account that the condensate is always accompanied by thermal atoms, the model function is extended by an additional gaussian function $\rho_{\text{th}}(z)$ representing the density distribution of the thermal atoms. The width of the gaussian was kept constant for all profile analysis and estimated from the $d=0$ displacement images where no phase imprinting appears. The number of atoms in the thermal part was set to be a constant fraction of the total observed atom number. 

\begin{figure}[t]
\centerline{\scalebox{0.65}{\includegraphics{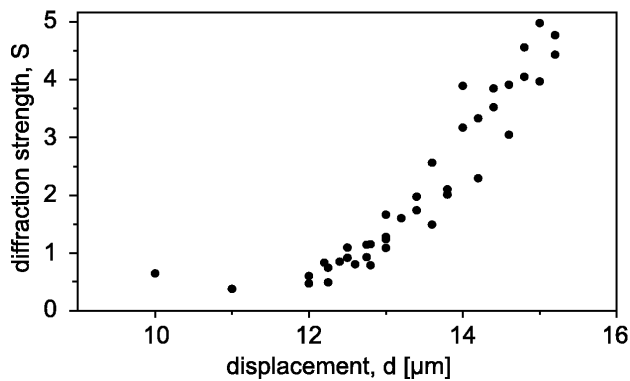}}}
\caption{For each displacement $d$ the fit of the density profiles (Fig. \ref{fig:Figure3}) yield the corresponding imprint parameter $S$. Significant phase imprinting starts at a displacement of about $12\mu\text{m}$ and raises steadily with increasing displacement.} \label{fig:Figure4}
\end{figure}

The experimental data is analysed by fitting the density profiles with the model function. Fig. \ref{fig:Figure3} shows the resulting fit (solid line) for three different displacements. The contributions of the different diffraction orders and the thermal component are plotted as solid and dashed lines, respectively. The overall envelope of the density profile is well described by the model. Additional structure (e.g. the narrow double peak for $d=13\mu\text{m}$) may arise due to interference of the different diffraction orders, which was not taken into account in the analysis. For each absorption image the fit yields a value for the imprint parameter $S$ which is shown as a function of the displacement $d$ in Figure \ref{fig:Figure4}. Diffraction becomes apparent for $d>12\mu\text{m}$ and the imprint parameter increases steadily with $d$. 

The interaction time $t_{\text{int}}$ with the lattice can be calculated from the imprint parameter $S=U t_{\text{int}}/\hbar$, with U being the amplitude of the potential modulation at the lattice. An upper limit for $t_{\text{int}}$ can be estimated by taking $U=(U_{\text{max}}-U_{\text{sad}})/2$ where $U_{\text{max}}=1/2\cdot m\omega^2 d^2$ is the potential energy after the displacement and $U_{\text{sad}}$ the potential energy at the saddle point. For the maximum displacement of $d=15.2\mu\text{m}$ $(U_{\text{max}}-U_{\text{sad}})/2\hbar=1.2\cdot 10^4 s^{-1}$, such that the observed maximum value of $S=5$ is reached within 0.4 ms. This corresponds to about 3\% of the total vertical oscillation period of 13.2 ms.  The atoms in the first diffraction order cover during this time an axial distance of $v_l t_{\text{int}}=0.46\mu\text{m}$. This is significantly smaller than the lattice constant of $4\mu\text{m}$ which is consistent with the Raman Nath approximation.

The experiment shows that diffraction from a magnetic lattice is possible with tolerable losses. This opens up novel perspectives for integrated matter wave interferometers. By allowing the condensate to interact twice with the lattice during two full vertical oscillations, a temporal interferometer may be realized. It would be sensitive to any phase change acquired by the condensate during the time between the two interactions with the lattice. A force applied along the axial direction could be detected in the change of the interference pattern. This would allow for a sensitive interferometric force detector integrated on a chip.

In summary we have demonstrated diffraction of Bose-Einstein condensates from a periodic potential generated by miniaturized current conductors on a micro chip. This realization of matter wave interference in a pure magnetic micro trap could open the door for future integrated atomic mater wave optics on a micro chip.

We thank C. Vale for experimental contributions and P. Schlagheck for valuable discussions. We acknowledge support by the Deutsche Forschungsgemeinschaft, the Landesstiftung Baden W\"urttemberg and the European Union.


\begin{thebibliography}{20}
\bibitem{comment1}
\bibinfo{author}{\bibnamefont{The current scheme is realized by high quality 300nm thick gold conductors on a $250\mu\text{m}$ thick silicon substrate. The conductors are fabricated using gold evaporation, electron-beam lithography, and dry-etching. They can carry a maximum current density of $5\cdot 10^6 \text{A/cm}^2$.}}
\bibitem{comment2}
\bibinfo{author}{\bibnamefont{The spatial calibration is based on the numerical simulation of the magnetic field for the well defined conductor geometry. It is cross checked by observing the loss of a condensate touching the chip surface.}}
\end{thebibliography}

\end{document}